\documentclass[article, twocolumn]{revtex4}
\usepackage[english]{babel}
\usepackage{amssymb,amsmath}
\usepackage{graphicx}
\usepackage{multirow}
\usepackage{bigstrut}

\begin{document}

\title{Bilayer manganites: polarons in the midst of a metallic breakdown}

\author{F. Massee \footnote{These authors contributed equally to this work}}\affiliation{Van der Waals-Zeeman Institute, University of Amsterdam, 1018XE Amsterdam, The Netherlands}
\author{S. de Jong*}\affiliation{Van der Waals-Zeeman Institute, University of Amsterdam, 1018XE Amsterdam, The Netherlands}
\author{Y. Huang}\affiliation{Van der Waals-Zeeman Institute, University of Amsterdam, 1018XE Amsterdam, The Netherlands}
\author{W.K. Siu}\affiliation{Van der Waals-Zeeman Institute, University of Amsterdam, 1018XE Amsterdam, The Netherlands}
\author{I. Santoso}\affiliation{Van der Waals-Zeeman Institute, University of Amsterdam, 1018XE Amsterdam, The Netherlands}
\author{A. Mans}\affiliation{Van der Waals-Zeeman Institute, University of Amsterdam, 1018XE Amsterdam, The Netherlands}
\author{A. T. Boothroyd}\affiliation{Clarendon Laboratory, Oxford University, OX1 3PU Oxford (UK)}
\author{D. Prabhakaran}\affiliation{Clarendon Laboratory, Oxford University, OX1 3PU Oxford (UK)}
\author{R. Follath}\affiliation{BESSY GmbH, Albert-Einstein-Strasse 15, 12489 Berlin, Germany}
\author{A. Varykhalov}\affiliation{BESSY GmbH, Albert-Einstein-Strasse 15, 12489 Berlin, Germany}
\author{L. Patthey}\affiliation{Paul Scherrer Institute, Swiss Light Source, CH-5232 Villigen, Switzerland}
\author{M. Shi}\affiliation{Paul Scherrer Institute, Swiss Light Source, CH-5232 Villigen, Switzerland}
\author{J.B. Goedkoop}\affiliation{Van der Waals-Zeeman Institute, University of Amsterdam, 1018XE Amsterdam, The Netherlands}
\author{M.S. Golden}\affiliation{Van der Waals-Zeeman Institute, University of Amsterdam, 1018XE Amsterdam, The Netherlands}\email{M.S.Golden@uva.nl}

\date{\today}

\maketitle

\textbf{The exact nature of the low temperature electronic phase of the manganite materials family, and hence the origin of their colossal magnetoresistant (CMR) effect, is still under heavy debate. By combining new photoemission and tunneling data, we show that in La$_{\text{2-2x}}$Sr$_{\text{1+2x}}$Mn$_2$O$_7$ the polaronic degrees of freedom win out across the CMR region of the phase diagram. This means that the generic ground state is that of a system in which strong electron-lattice interactions result in vanishing coherent quasi--particle spectral weight at the Fermi level for all locations in $k$--space. The incoherence of the charge carriers offers a unifying explanation for the anomalous charge-carrier dynamics seen in transport, optics and electron spectroscopic data. The stacking number $N$ is the key factor for true metallic behavior, as an intergrowth-driven breakdown of the polaronic domination to give a metal possessing a traditional Fermi surface is seen in the bilayer system.}



Competition between local lattice distortions leading to anti-ferromagnetic, charge and orbital (CO) ordering on the one hand, and mixed valence character promoting metallic ferromagnetic double exchange on the other, determines the transport transport properties \cite{Jin} of the manganite materials family and is proposed to lie at the root of their CMR effect \cite{SenPRL2007}. (La,Sr)$_{\text{N+1}}$Mn$_{\text{N}}$O$_{\text{3N+1}}$, abbreviated LSMO (where $N$ is the number of stacked MnO$_2$ planes between [La,Sr]O block layers \cite{goodenough}) displays a remarkable decrease in metallic character with decreasing $N$ \cite{kimura}, see Fig.~S1. Metallic behavior thrives in the phase diagram of cubic LSMO \cite{Urushibara}, whereas the bilayer analogue is metallic only in a narrow Sr--doping and temperature regime \cite{Moritomo}, giving rise to the largest CMR effect \cite{Perring}. The more strongly 2D, single layer compound shows neither metallic nor CMR behavior \cite{Moritomo_singlelayernoCMR}.

Focusing on bilayer LSMO within the CMR-region of the phase diagram, the prevailing picture from structural studies is one of polarons existing above T$_{\text{C}}$. On cooling towards T$_{\text{C}}$, these short range versions of the CO order typical of the insulating compositions, become increasingly correlated \cite{Vasiliu_polaron_correlations,Campbell}. Eventually double exchange, leading to an itinerant, metallic state, takes over. 

This metallic state for $x=0.4$ has been shown to support small quasi--particles (QPs) in the spectral function measured by angle-resolved photoemission (ARPES) \cite{mannella}. These signal coherent electronic excitations, albeit strongly dressed with lattice distortions, and are seen as evidence for a novel and elusive state of matter known as a polaronic metal \cite{mannella, mannella2}. 

Other ARPES studies paint a different picture, with stronger QP features observed at low T that persist up to temperatures of order 1.5T$_{\text{C}}$ \cite{dessau, dessau2, dejong}, despite the system being nominally insulating. In contrast, scanning tunneling microscopy / spectroscopy (STM/S) studies reported gaps in the local density of states near--E$_{\text{F}}$ for $x=0.30$ \cite{ronnow} and 0.325 \cite{desantis}, both in the metallic and insulating temperature regimes. Finally, new neutron diffraction data for $x=0.4$ has shown that even far below T$_{\text{C}}$ - at 10 K in the metallic state - polarons remain as fluctuations that strongly broaden and soften phonons near the wave vectors where the charge order peaks would appear in the insulating phase \cite{weber}. 

\begin{figure*}[!tb]
\includegraphics[width=2.0\columnwidth]{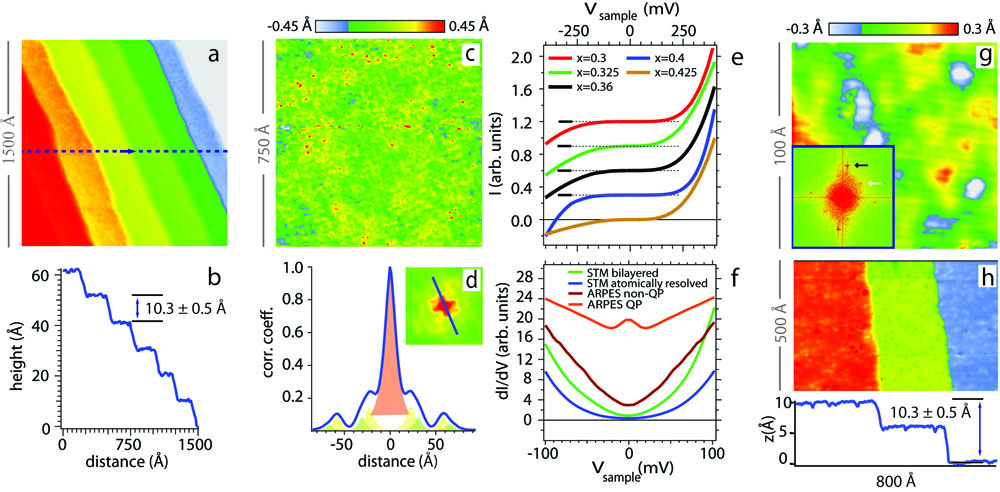}
\caption{\label{stm} STM/S on bilayer manganites. (a) Large field of view of a region with (an above average number of) half unit cell steps at the surface and (b) corresponding line scan ($x=0.3$, $V=-500$~meV, $I=20$~pA). (c) Semi-ordered square-like texture for x=0.36, ($V=-450$~meV, $I=150$~pA), which is more clearly seen in the line trace (d) through the center of the autocorrelation (inset of d, 220$\times$220\AA$^{2}$). A correlation length of $\sim$2.3 nm is determined from fitting Lorentzians to the autocorrelation features. (e) STS spectra - I(V) curves - showing a gap of more than 100 meV for all doping concentrations studied ($V=500$~meV, $I\approx 50$~pA). The spectra offset with respect to each other for clarity, indicated by the dashed lines. (f) Comparison between symmetrised, k--integrated ARPES spectra and STM dI/dV traces. (g) Topograph displaying atomic resolution ($V=500$~meV, $I=140$~pA) with (1$\times$1) and $\sqrt{2}\times\sqrt{2}$ spots in the Fourier transform (see inset, black and white arrows respectively) which are not found in the corresponding low energy electron diffraction images. (h) Topography of stepped surface taken on the same region as (g), where the step height corresponds to that expected for single layer LSMO.}
\end{figure*}

Here, a combination of ARPES and STM/S reveals that the bilayer manganites still have a number of tricks up their sleeves. Firstly, we show that the intrinsic spectral response of these systems is pseudogapped, with negligible coherent spectral weight at E$_{\text{F}}$ anywhere in k--space at any temperature, across the CMR-region of the phase diagram. Secondly, we show that the strong QP features seen in ARPES studies - also above T$_{\text{C}}$ - are due to the unavoidable presence of $N\neq2$ stacking-fault intergrowths. 

These new insights clear the way for a unified interpretation of the physical properties of CMR bilayer LSMO in terms of strongly incoherent charge carriers, and suggest that local control of the dimensionality of such manganites - via tuning the stacking number, N - may offer a novel route to new functional nanostructures.

Imaging the cleavage surface of a typical bilayer LSMO single crystal in the CMR doping region using low temperature STM/S yields large, flat terraces, see Figs.~1a and b. One surprising characteristic of these STM data is that the surface atoms have proven extremely difficult to image. In the literature, the only reported observations of atomic resolution in bilayer LSMO concerned small nm-sized patches \cite{ronnow, desantis}. The terraced, flat and debris-free surfaces we image may lack atomic corrugation, however they do posses a spatial texture in the tunneling signal. These structures are usually disordered, but are - in some cases - ordered into a semi-regular, square-like lattice, displaying characteristic length scales of order 2-3~nm, as can be seen in Fig.~1c. The quasi-ordered nature of these regions at the surface is evident from the autocorrelation traces from the STM topographs, as shown in Fig.~1d, which display clear structures at distances between 5-15 units cells. 

Comparison with low energy electron diffraction (LEED) data recorded in the STM chamber shows that the orientation of the quasi-periodic structures is not pinned to the underlying atomic lattice. The tunneling spectra of bilayer LSMO from all cleavage surfaces measured are gapped (symmetrically) around E$_{\text{F}}$ over an energy range of several 100~meV at low temperature for all doping concentrations studied, as shown in Fig.~1e.  
The tunneling spectra themselves show no major variations from point to point. This suggests that the observed spatial textures seen in the STM data are a subtle, `higher order' structural ordering phenomenon, leading to modest, periodic signatures, somewhat reminiscent of the 2.5 nm correlated regions seen in the high-T, polaronic liquid phase described in Ref.~\cite{Campbell}. The authors of Ref.~\cite{SenPRL2007} suggest that aggregates of polarons resembling a charge ordered state should be observable in STM experiments above T$_{\text{C}}$. Our new data show that these kind of entities could exist at the surface even at low temperatures, possibly being a log-jammed form of the fluctuating polarons observed in recent neutron experiments \cite{weber}.

During our STM investigation of numerous cleavage surfaces of bilayer LSMO, very clear atomic resolution was found over a large terrace of one particular cleave, (shown in Fig.~1g). Analysis of the Fourier transform shows reduced in-plane symmetry, with clear $\sqrt{2}\times\sqrt{2}$ spots shown in the inset to Fig.~1g. This is atypical for bilayer LSMO, which has a $(1\times1)$ tetragonal lattice symmetry in the ferromagnetic `metallic' phase that we observe in all our LEED data (see Fig.~S2c). An important observation linked to this large scale STM imaging with atomic contrast is that the step bordering this terrace is only one quarter of the $N=2$ c-axis unit cell in height (5 \AA, see Fig.~1h). Additionally, the STS spectra from this region were more strongly gapped compared to those from non-atomically resolved regions displaying 10 \AA\ step heights, as shown in Fig.~1f. Taken together, these facts form a compelling argument that in Figs.~1g and h we are - in fact - imaging a stacking fault in the Ruddlesden-Popper manganite in which an extra La$_{2}$O$_{2}$ block (or blocks) creates a region of the bilayer crystal which is effectively single layer (La,Sr)$_{2}$MnO$_{4}$. The fact that atomic contrast is readily observed in STM of $N=1$ LSMO \cite{Evtushinsky_single_layer_STM} also further strengthens our interpretation.

The observation with STM/S of such a surface inclusion with a different stacking number $N$ is rare, but is not, in itself, wholly surprising. Numerous $\mu$SR studies \cite{Allodi_intergrowths_muSR}, magnetization measurements \cite{Potter_intergrowths_magnetisation1, Bader_intergrowths_magn_TEM} and transmission electron microscopy studies \cite{Seshadri_intergrowths_TEM1, Sloan_intergrowths_TEM2} have established that stacking faults occur even in the very best crystals at the $\approx1$\% level \cite{Potter_intergrowths_magnetisation1, Bader_intergrowths_magn_TEM}. These intergrowths vary - locally - the stacking number $N$ and are the cause of anomalous steps in the magnetization above the bulk T$_{\text{C}}$ common in bilayer LSMO between 200 and 350~K: small patches with $N$ values above two deliver a higher T$_{\text{C}}$. Due to the strong connection between magnetism and metallic behavior in the manganites, it is a simple step to reason that $N>2$ intergrowths would also be metallic above and beyond the bilayer T$_{\text{C}}$. In contrast, an $N=1$ intergrowth will be more insulating than bilayer LSMO, in keeping with the STM/S data shown in Fig.~1f.

Armed with a heightened awareness that (unavoidable) intergrowths in bilayer LSMO are present in real samples, and that signals from these regions can be picked up in spectroscopic experiments, we now turn to our ARPES investigations. In total, we have measured dozens of bilayer LSMO single crystals from two different sources with Sr doping levels spanning the entire CMR region of the phase diagram. ARPES spectra were taken at low temperature over a square grid across the samples in steps of 100 $\mu$m, see Fig.~2b. Using such a procedure, two types of spectral signature are observed on different locations: (i) from only $1-5$\% of the cleavage surface, sharp QP--peaks are observed at all $k$--locations on the Fermi surface, as shown in Fig.~2c and (ii) vanishingly small spectral weight is observed near E$_{\text{F}}$ from the majority of the sample surface, as can be seen in Fig.~2a for 10 K and in the comparison made in Fig.~2d. Fig.~2 also shows the ARPES signatures from the non--QP--peaked and QP--peaked regions for T$>$T$_{\text{C}}$. For the non--QP--peaked regions, a modest shift in spectral weight to higher binding energies can be seen. For the minority, QP--peaked locations, the near E$_{\text{F}}$ spectral weight is reduced at 150K (a temperature that is almost twice the bilayer T$_{\text{C}}$ for this doping), however the $I(E)$--traces in Fig.~2d still display considerable intensity at E$_{\text{F}}$, and certainly no signs of a pseudogap. This despite the fact that the vast majority of the sample is in the paramagnetic, insulating regime. In Figs.~S3 and S4, we show that for varying doping levels the Fermi surface (FS) topology (i.e. the number of bands), the FS area and the QP--intensity vs. temperature for the QP--peaked regions are not in line with the bulk properties of bilayer LSMO. Additionally, when observed for a particular doping, the QP--peaks display a cleave-to-cleave variation in the number of FS sheets observed around the X--point ranging from one to four.

\begin{figure*}[!tb]
\includegraphics[width=2.0\columnwidth]{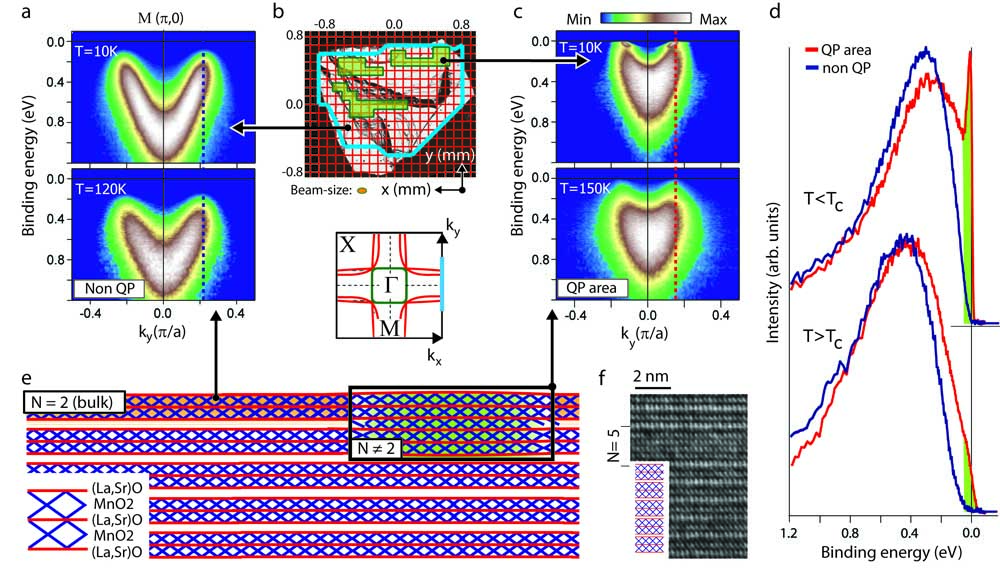}
\caption{\label{arpes} Spatial ARPES mapping of bilayer LSMO. (a) Typical $I(E,k)$--images taken around the $(\pi,0)$ point of $k$--space for temperatures well below and above T$_{\text{C}}$ for an $x=0.425$ sample from a non--QP--peaked region of the cleave (h$\nu$ = 56 eV). A clear absence of QP features characterizes the data both above and below the bulk T$_{\text{C}}$. (b) Optical micrograph of the post-cleavage surface of a bilayer LSMO crystal (from an $x=0.36$ sample). At each red grid point within the blue box an ARPES spectrum has been taken. The synchrotron spot size is shown to scale. The regions which do support QP--peaked spectra are indicated by the green shaded areas. The proportion of the surface showing QP--peaked spectra is unusually abundant for this sample. The inset below panel b shows a schematic of the Brillouin zone, after Ref.~\cite{LDA}, with the ARPES cuts shown in a and c indicated in blue. (c) Analogous data to those of panel a, now for a QP--peaked region of the same sample. (d) $I(E)$--curves taken along the dashed lines in (a) and (c). For the QP--peaked region, there remains significant spectral weight at E$_F$ for 150K, despite the fact that T $\rightarrow$ 2 T$_{\text{C}}$. In sharp contrast, the non--QP--peaked $I(E)$ curves display a pseudogap-like lack of E$_F$--spectral weight both below and above T$_{\text{C}}$. Panel (e) shows the explanation for the striking lateral position dependence of the electronic structure. Depicted is a schematic side-on view of a cleaved bilayer LSMO crystal with a stacking fault (illustrated is an $N=5$ intergrowth) just under the crystal surface. Such inclusions support QP--states to temperatures well above the bilayer T$_{\text{C}}$. The \emph{intrinsic} spectral signature of bilayer LSMO is that seen in panel a and the blue traces in panel d. (f) In connection to the schematic in e: a transmission electron microscopy image, adapted from \cite{Bader_intergrowths_magn_TEM}, showing an $N=5$ inclusion in a bilayer LSMO single crystal.}
\end{figure*}

Thus, combining our atomic scale tunneling data with these ARPES data it is clear that the strongly QP--peaked spectral signature is \emph{not} representative of bilayer LSMO. Instead, the intrinsic signature of this material is that of a pseudo--gapped, very bad metal that supports vanishingly small QP spectral weight at E$_{\text{F}}$ at any $k$--location. This may sound surprising for a metal, but the bilayer manganites are far from being normal metals. In the `metallic' phase, the ab-plane resistivity is in excess of the Mott maximum of $10^{-3}$~$\Omega$cm \cite{Chudnovskii_Mott_limit_1978, Mott_Mott_limit_1974}, which is quite unlike the situation in their cubic cousins (see Fig.~S1). 
In addition, the optical conductivity of bilayer LSMO shows an incoherent Drude peak, even down to the lowest temperatures \cite{Ishikawa_optics_lsmo1, Takahashi_optics_lsmo2}, again, unlike the situation in the $N=\infty$, cubic materials \cite{Okimoto_optics_cubic}. Therefore, it is clear that the strongly QP--peaked regions of the cleaves are simply signals from stacking faults \cite{Bader_intergrowths_magn_TEM} located sufficiently close to the crystal termination (Fig.~2e and 2f) which consequently contribute strongly to the near-E$_{\text{F}}$ photoemission signal. In this manner, the anomalous temperature dependence of the QP--peaks (Fig.~S4c) also falls naturally into place. Fig.~S5 provides additional arguments involving the orthorhombic crystal symmetry of the stacking fault regions observed in ARPES data; here we note that the inset to Fig.~1g already shows the orthorhombic surface of an intergrowth (in this case $N=1$).

The new experimental data from real space and $k$--space probes presented above provides a unifying framework in which to understand all the published ARPES data, which is a welcome simplification. However, the bilayer manganites still present a richness - in particular in the form of a highly sensitive doping dependence - which is a major challenge to our theoretical understanding. Although in our experiments we have adopted a wide range of experimental conditions including geometries and polarizations \cite{footnote_pol} very similar to those of Refs.~\cite{mannella} and \cite{mannella2}, we have not observed the nodal quasiparticles reported for the $x=0.4$ composition. This may seem puzzling until one realizes that the phase diagram of bilayer LSMO possesses numerous line phases. Various narrow regions in the phase diagram show special types of magnetic and/or orbital ordering: $x=0.30$ (AFM metal), 0.50 (CE--type charge and orbital ordering) and 0.60 (AFM metal) \cite{phase_0.3, linephase_0.5, linephase_0.6}. Remarkably, deviations of only 0.01 in doping level (e.g. away from $x=0.60$ [\onlinecite{linephase_0.6,Sun_0.6}]) have a major effect on the electronic behavior. Accordingly, we suggest that exactly at or very close to a composition of $x=0.40$ there is also such a line phase, whose exact composition has not been `hit' in the ARPES data of Refs.~\cite{Chuang_arpes_gapped1, Kubota_arpes_gapped2,dessau, dessau2, dejong}.

Our data - covering samples right across the CMR region of the phase diagram - shows that the generic electronic signature of bilayer LSMO is that of a pseudo--gapped, very bad metal with vanishingly small QP spectral weight at the Fermi level. Hence, `nodal metallicity', although of great interest in the context of comparisons with the high T$_{\text{c}}$ superconducting cuprates, is apparently not a necessary factor for bringing about CMR behavior, and appears to be confined to the doping level $x=0.40$.

Before reaching a final conclusion as regards the physics behind a metal which has QP spectral weight below our detection limit, one further issue needs to be dealt with, and that is whether the surface of bilayer LSMO is even less like a regular metal than the bulk. X-ray resonant diffraction has shown the first pair of MnO$_2$--planes nearest to the surface of bilayer LSMO cleaved in air \cite{surface_bulk, freeland_nature} or in vacuum \cite{nascimento} to possess no ferromagnetic order and thus the surface could be expected to display non-metallic behavior. A recent (hard) x-ray photoemission study of the same crystals as studied here found no major difference between the surface and bulk in terms of charge transfer or composition \cite{x-ray_manganites}, thus the differences between the outermost bilayer and the rest are subtle in nature. While recent LEED studies \cite{nascimento} have given no evidence for a lowering of the 2D surface symmetry, for example via a surface reconstruction, a finding our own LEED data support, a contraction of the apical bond length for the outermost MnO$_2$-plane was observed \cite{nascimento}, which could impact the mobility of the charge carriers at the surface. 

Here it is essential to realize that the photocurrent measured in ARPES certainly originates from deeper in the crystal than the first bilayer only, even at excitation energies in the vacuum ultraviolet. Therefore, we can conclude that the bad-metallic, fully pseudo-gapped behavior we observe is indeed a genuine characteristic of the bulk for the vast majority of the CMR-compositions in bilayer LSMO. Although disentangling the spectral signatures of the outermost and deeper lying bilayers is difficult, we do note that the energy range at E$_{\text{F}}$ over which the spectral weight is suppressed in ARPES is smaller than the negative bias gap seen in STM/S, the latter probing strictly the electronic states at the surface. This is evident from Fig.~1f where STS spectra from bilayer LSMO and from the $N=1$ intergrowth are compared with symmetrized, k--integrated photoemission spectra from ARPES maps (such as those in Fig.~S3) for a QP--peaked and a non--QP--peaked region of the surface.

We now return to the key issue of the physical nature of the charge carriers responsible for electronic transport of the bilayer manganites. The most suitable picture - considering the results on both the dynamics of the charge carriers and the subtle yet clear structural anomalies they bring with them - is that of polaronic charge carriers above T$_{\text{C}}$, but also in the ferromagnetic (bad) metallic state. The fact that the STM signatures of the weakly self-organised polaronic carriers when `log-jammed' at the surface resemble the polaronic correlations seen above T$_{\text{C}}$ suggest that in the bulk of the bilayer LSMO systems - which is more metallic than the surface - the system teeters on the edge of a breakdown of the fragile polaronic metal state into an insulting, charge and orbitally ordered state. This borderline situation, coupled to the disorder and enhanced fluctuations present in these quasi-2D systems, delivers all the ingredients for the colossal magnetoresistance transition \cite{SenPRL2007}.    

We conclude this paper by making a connection back to the relationship between the materials physics and crystal chemistry of the manganite family. The physics describing the manganites depends crucially on the propensity of the system to form ordered textures in spin, charge and orbital occupation. For the polaronic metal state to remain stable, it is vital that the degeneracy in the $e_g$ orbital manifold is preserved, as this is the ticket to the double-exchange energy reduction that encourages hopping of the carriers. This condition - in turn - stipulates equality in the equatorial and axial bond lengths of the MnO$_6$ octahedra. The deviating axial bond lengths at the surface of bilayered LSMO compared to the bulk, as observed in Ref.~\cite{nascimento} lift this degeneracy and thus push the surface of this material further into the insulating regime than bulk: the correlated polarons can become static, as we observe with our STM measurements. On the other hand, the $N>2$ inclusions we observe, structurally bear greater resemblance to the cubic compound, and thus show a higher propensity towards metallic behavior, including the existence of coherent spectral weight. Paradoxically, the fragility of the polaronic metal state in the bilayer systems is also the key to their colossal magnetoresistance, as it delivers the precarious balance between weakly metallic and insulating behavior required for such an enormous sensitivity to the extra impulse provided by an external magnetic field. 

The results presented here formulate a clear challenge to develop a general theory for the transport in $N=2$ systems involving practically incoherent charge carriers - fluctuating polarons - while also capturing the sensitivity to the stacking number, $N$. Aside from the lattice polaron generally considered, different types of polaron, such as spin and orbital polarons, should be taken into consideration \cite{vandenbrink_prl_85, bala_prl_87, bala_prb_65}. An interesting proposal for the charge carrier dynamics in the manganites is the existence of so called Zener polarons \cite{aladine_prl_89, aladine_prl_101}. In such a polaron, the charge carrier is not localised on a single manganese atom, but on two neighbouring manganese atoms which are ferromagnetically coupled by the Zener double-exchange mechanism. It has been shown that all these different types of polaron lead to a large incoherent spectral weight such as is seen in our ARPES investigations \cite{vandenbrink_prl_85, bala_prl_87, bala_prb_65, wohlfeld_prb_79}.

The new insight we have gained points towards the great potential of e.g. layer-by-layer thin--film engineering to generate tailor-made heterostructures, not only to lead to enhanced transition temperatures \cite{May_LMO_SMO_multi}, but in combination with modern lithographic and patterning methods to tune and improve magnetoresistive properties on the sub--micron scale in a new generation of complex oxide devices.

%

\section{Methods.}

\subparagraph{STM.}
The STM data were recorded using a commercial variable temperature UHV microscope from Createc GmbH. W tips were prepared by electrochemical etching followed by in-situ conditioning before each measurement using a Au(788) single crystal. In all cases, the spectral shapes obtained were independent of the tip to sample distance, a sign of good vacuum tunneling conditions. All investigated samples have been cleaved in situ prior to measuring. The crystallographic orientation was determined directly after the STM measurements using in situ low energy electron diffraction. The setup voltages, as well as the tunneling currents are indicated in the figures.

\subparagraph{ARPES.}
The angle-resolved photoemission data presented here were obtained using the following beamlines and end-stations: i) the UE112-PGMa beamline at the Helmholtz Zentrum Berlin (BESSY II storage ring), Berlin, Germany, coupled to an SES100 analyzer; ii) the UE112-PGMb beamline at BESSY coupled to the R4000 analyzer of the 1cubed end-station and iii) the SIS beamline at the Swiss Light Source (SLS), Villigen, Switzerland, equipped with an SES2002 analyzer. The total experimental energy broadening at 20 K was set to 30~meV, 20~meV and 15~meV, for the three end-stations, respectively. The momentum resolution was 0.02$\pi$/a along the analyzer slit at the excitation energies used. The photon energies used are indicated in the relevant figures. High quality single crystals of LSMO were grown in both Oxford and Amsterdam using optical floating zone techniques, and were cleaved in situ and characterized using LEED before ARPES measurements were conducted. For further details concerning the sample quality and characterization, see Fig.~S2.

\begin{acknowledgments}
We thank R. Huisman, R. Zeiler and M. Gobbi for help during the beamtimes, F.D. Tichelaar and H. Zandbergen for TEM investigations, H. Luigjes, H. Schlatter and J. S. Agema for valuable technical support and the IFW Dresden ARPES group for access to the SES100 end station. We are grateful to N. Mannella, G. A. Sawatzky, A.J. Millis, P. Littlewood, E. van Heumen and J. Zaanen for useful discussions. This work is part of the research program of the FOM, which is financially supported by the NWO.
\end{acknowledgments}


\begin{thebibliography}{99}

\bibitem{Jin}
S. Jin, T. H. Tiefel, M. McCormack, R. A. Fastnacht, R. Ramesh
and L. H. Chen, Science \textbf{264}, 413 (1994)

\bibitem{SenPRL2007}
Cengiz Sen, Gonzalo Alvarez and Elbio Dagotto, Phys. Rev. Lett. \textbf{98}, 127202 (2007)

\bibitem{goodenough}
J. B. Goodenough, Phys. Rev. \textbf{100}, 564 (1955)

\bibitem{kimura}
T. Kimura and Y. Tokura, Annu. Rev. Matter. Sci. \textbf{30}, 451 (2000)

\bibitem{Urushibara}
 A. Urushibara, Y. Moritomo, T. Arima, A. Asamitsu, G. Kido and Y. Tokura, Phys. Rev. B \textbf{51}, 14103 (1995)

\bibitem{Moritomo}
Y. Moritomo, A. Asamitsu, H. Kuwahara and Y. Tokura, Nature (London) \textbf{380}, 141 (1996) 

\bibitem{Perring}
T. G. Perring, G. Aeppli, Y. Moritomo and Y. Tokura, Phys. Rev. Lett. \textbf{16}, 3197 (1997)

\bibitem{Moritomo_singlelayernoCMR}
Y. Moritomo, Y. Tomioka, A. Asamitsu, Y. Tokura and Y. Matsui , Phys. Rev. B \textbf{51}, 3297 (1995)

\bibitem{Vasiliu_polaron_correlations}
L. Vasiliu-Doloc, S. Rosenkranz, R. Osborn, S. K. Sinha, J. W. Lynn, J. Mesot, O. H. Seeck, G. Preosti, A. J. Fedro and J. F. Mitchell, Phys. Rev. Lett. \textbf{83}, 4393 (1999)

\bibitem{Campbell}
B. J. Campbell,R. Osborn, D. N. Argyriou,L. Vasiliu-Doloc, J. F. Mitchell, S. K. Sinha, U. Ruett, C. D. Ling, Z. Islam, and J. W. Lynn, Phys. Rev. B  \textbf{65}, 014427 (2001)

\bibitem{mannella}
N. Mannella, W. L. Yang, X. J. Zhou, H. Zheng, J. F. Mitchell, J. Zaanen, T. P.
Devereaux, N. Nagaosa, Z. Hussain and Z.-X. Shen, Nature \textbf{438}, 474 (2005)

\bibitem{mannella2}
N. Mannella, W. L. Yang, X. J. Zhou, H. Zheng, J. F. Mitchell, J. Zaanen, T. P.
Devereaux, N. Nagaosa, Z. Hussain, and Z.-X. Shen, Phys.
Rev. B \textbf{76}, 233102 (2007)

\bibitem{dessau}
Z. Sun, Y.-D. Chuang, A. V. Fedorov, J. F. Douglas, D. Reznik, F. Weber, N. Aliouane, D. N. Argyriou, H. Zheng, J. F. Mitchell, T. Kimura, Y. Tokura, A. Revcolevschi and D. S. Dessau. Phys. Rev. Lett. \textbf{97}, 056401 (2006)

\bibitem{dessau2}
Z. Sun, J. F. Douglas, A. V. Fedorov, Y.-D. Chuang, H. Zheng, J. F. Mitchell and D. S. Dessau. Nature Physics \textbf{3}, 248 (2007)

\bibitem{dejong}
S. de Jong, Y. Huang, I. Santoso, F. Massee, R. Follath, O. Schwarzkopf, L. Patthey, M. Shi, and M. S. Golden, Phys. Rev. B \textbf{76}, 235117 (2007)

\bibitem{ronnow}
H. M. R{\o}nnow, Ch. Renner, G. Aeppli, T. Kimura and Y. Tokura, Nature \textbf{440}, 1025 (2006)

\bibitem{desantis}
S. De Santis, B. Bryant, M. Warner, H. Wang, T. Kimura, Y, Tokura,	Ch. Renner, A. Bianconi and G. Aeppli, J. Supercond. Nov. Magn. \textbf{20}, 531 (2007)

\bibitem{weber}
F. Weber, N. Aliouane, H. Zheng, J. F. Mitchell, D. N. Argyriou and D. Reznik, Nature Materials \textbf{8}, 798 (2009)

\bibitem{Evtushinsky_single_layer_STM}
D. V. Evtushinsky, D. S. Inosov, G. Urbanik, V. B. Zabolotnyy, R. Schuster, P. Sass, T. H\"{a}nke,
C. Hess, B. B\"{u}chner, R. Follath, P. Reutler, A. Revcolevschi, A. A. Kordyuk and S. V. Borisenko, Phys. Rev. Lett. \textbf{105}, 147201 (2010)

\bibitem{Allodi_intergrowths_muSR}
G. Allodi, M. Bimbi, R. D. Renzi, C. Baumann, M. Apostu, R. Suryanarayanan
and A. Revcolevschi, Phys. Rev. B \textbf{78}, 064420 (2008)

\bibitem{Potter_intergrowths_magnetisation1}
C. D. Potter, M. Swiatek, S. D. Bader, D. N. Argyriou, J. F. Mitchell, D. J.
Miller, D. G. Hinks and J. D. Jorgensen, Phys. Rev. B \textbf{57}, 72 (1998)

\bibitem{Bader_intergrowths_magn_TEM}
S. D. Bader, R. M. Osgood, D. J. Miller, J. F. Mitchell and J. S. Jiang, J.
Appl. Phys. \textbf{83}, 6385 (1998)

\bibitem{Seshadri_intergrowths_TEM1}
R. Seshadri, M. Hervieu, C. Martin, A. Maignan, B. Domenges, B. Raveau and
A. N. Fitch, Chem. Mater. \textbf{9}, 1778 (1997)

\bibitem{Sloan_intergrowths_TEM2}
J. Sloan, P. D. Battle, M. A. Green, M. J. Rosseinsky and J. F. Vente, J. Sol. State Chem. \textbf{138}, 135 (1998)

\bibitem{LDA}
X. Y. Huang, O. N. Mryasov, D. L. Novikov and A. J. Freeman, Phys. Rev. B \textbf{62}, 13318 (2000)

\bibitem{Chudnovskii_Mott_limit_1978}
F. A. Chudnovskii, J. Phys. C: Solid State Phys., \textbf{11}, L99 (1978)

\bibitem{Mott_Mott_limit_1974}
N. F. Mott, Metal-Insulator Transitions. London: Taylor and Francis (1974)

\bibitem{Ishikawa_optics_lsmo1}
T. Ishikawa, K. Tobe, T. Kimura, T. Katsufuji and Y. Tokura, Physical Review B \textbf{62},12354 (2000)

\bibitem{Takahashi_optics_lsmo2}
K. Takahashi, N. Kida and M. Tonouchi, J. Mag. Mag. Mater. \textbf{272}, 669 (2004)

\bibitem{Okimoto_optics_cubic}
Y. Okimoto, T. Katsufuji, T. Ishikawa, T. Arima, and Y. Tokura, Phys. Rev. B \textbf{55}, 4206 (1997)

\bibitem{footnote_pol}
We have examined our crystals using ARPES employing different polarization conditions and measurement geometries. For all x=0.4 cleaves and for the non--QP--peaked regions of all other doping levels, use of a high degree of in-plane polarization (as related in Ref.~\cite{mannella}) did not uncover small QP--peaked structures close to E$_{\text{F}}$. Furthermore, as Fig.~S6 shows, the strong QP features from the $N\neq2$ intergrowths were robust with respect to wide variation of the polarization and geometry conditions.

\bibitem{phase_0.3}
T. Kimura, Y. Tomioka, H. Kuwahara, A. Asamitsu, M. Tamura and Y. Tokura, Science \textbf{274}, 1698 1996

\bibitem{linephase_0.5}
Q. A. Li, KE. Gray, H. Zheng, H. Claus, S. Rosenkranz, S. N. Ancona, R. Osborn, JF. Mitchell, Y. Chen and J. W. Lynn, Phys. Rev. Lett. \textbf{98}, 167201 (2007)

\bibitem{linephase_0.6}
H. Zheng, Q. Li, K. E. Gray and J. F. Mitchell, Phys. Rev. B \textbf{78}, 155103 (2008)

\bibitem{Sun_0.6}
Z. Sun, J. F. Douglas, Q. Wang, D. S. Dessau, A. V. Federov, H. Lin, S. Sahrakorpi, B. Barbiellini, R. S. Markiewicz, A. Bansil, H. Zheng and J.F. Mitchell, Phys. Rev. B \textbf{78}, 075101 (2008)

\bibitem{Chuang_arpes_gapped1}
Y.-D. Chuang, A. D. Gromko, D. S. Dessau, T. Kimura and Y. Tokura, Science \textbf{25}, 1509 (2001)

\bibitem{Kubota_arpes_gapped2}
M. Kubota, K. Onoa and T. Yoshida, J. Electron Spec. Rel. Phenom. \textbf{156}, 398 (2007)

\bibitem{surface_bulk}
J. W. Freeland, J. J. Kavich, K. E. Gray, L. Ozyuzer, H. Zheng, J. F. Mitchell, M. P. Warusawithana, P. Ryan, X. Zhai, R. H. Kodama and J. N. Eckstein, J. Phys.: Condens. Matter \textbf{19}, 315210 (2007)

\bibitem{freeland_nature}
J. W. Freeland, K. E. Gray, L. Ozyuzer, P. Berghuis, E. Badica, J. Kavich, H. Zheng and J. F. Mitchell, Nature Materials \textbf{4}, 63 (2005)

\bibitem{nascimento}
V. B. Nascimento, J.W. Freeland, R. Saniz, R. G. Moore, D. Mazur, H. Liu, M. H. Pan, J. Rundgren, K. E. Gray,
R. A. Rosenberg, H. Zheng, J. F. Mitchell, A. J. Freeman, K. Veltruska and E. W. Plummer, Phys. Rev. Lett. \textbf{103}, 227201 (2009)

\bibitem{x-ray_manganites}
S. de Jong, F. Massee, Y. Huang, M. Gorgoi, F. Schaefers, J. Fink, A. T. Boothroyd, D. Prabhakaran, J. B. Goedkoop and M. S. Golden, Phys. Rev. B \textbf{80}, 205108 (2009)

\bibitem{vandenbrink_prl_85}J. van den Brink, P. Horsch and A.~M. Ole\'{s}, Phys. Rev. Lett. \textbf{85}, 5174 (2000)
\bibitem{bala_prl_87}J. Ba{\l}a, G. A. Sawatzky, A. M. Ole\'{s} and A. Macridin, Phys. Rev. Lett. \textbf{87}, 067204 (2001)
\bibitem{bala_prb_65}J. Ba{\l}a, A. M. Ole\'{s} and P. Horsch, Phys. Rev. B \textbf{65}, 134420 (2002)

\bibitem{aladine_prl_89}A. Daoud-Aladine, J. Rodr\'{i}guez-Carvajal, L. Pinsard-Gaudart, M. T. Fern\'{a}ndez-D\'{i}az and A. Revcolevschi, Phys. Rev. Lett. \textbf{89}, 097205 (2002)
\bibitem{aladine_prl_101}A. Daoud-Aladine, C. Perca, L. Pinsard-Gaudart and J. Rodr\'{i}guez-Carvajal, Phys. Rev. Lett. \textbf{101}, 166404 (2008)
\bibitem{wohlfeld_prb_79}K. Wohlfeld, A. M. Ole\'{s} and P. Horsch, Phys. Rev. B \textbf{79}, 224433 (2009)

\bibitem{May_LMO_SMO_multi}
S. J. May, P. J. Ryan, J. L. Robertson, J.-W. Kim, T. S. Santos, E. Karapetrova, J. L. Zarestky, X. Zhai, S. G. E. te Velthuis, J. N. Eckstein, S. D. Bader and A. Bhattacharya, Nature Materials \textbf{8}, 892 (2009)


\end{thebibliography}
\end{document}


\title{Supplementary information to the article `Bilayered manganites: polarons in the midst of a metallic breakdown'}

\author{F. Massee*}\affiliation{Van der Waals-Zeeman Institute, University of Amsterdam, 1018XE Amsterdam, The Netherlands}
\author{S. de Jong*}\affiliation{Van der Waals-Zeeman Institute, University of Amsterdam, 1018XE Amsterdam, The Netherlands}
\author{Y. Huang}\affiliation{Van der Waals-Zeeman Institute, University of Amsterdam, 1018XE Amsterdam, The Netherlands}
\author{W.K. Siu}\affiliation{Van der Waals-Zeeman Institute, University of Amsterdam, 1018XE Amsterdam, The Netherlands}
\author{I. Santoso}\affiliation{Van der Waals-Zeeman Institute, University of Amsterdam, 1018XE Amsterdam, The Netherlands}
\author{A. Mans}\affiliation{Van der Waals-Zeeman Institute, University of Amsterdam, 1018XE Amsterdam, The Netherlands}
\author{A. T. Boothroyd}\affiliation{Clarendon Laboratory, Oxford University, OX1 3PU Oxford (UK)}
\author{D. Prabhakaran}\affiliation{Clarendon Laboratory, Oxford University, OX1 3PU Oxford (UK)}
\author{R. Follath}\affiliation{BESSY GmbH, Albert-Einstein-Strasse 15, 12489 Berlin, Germany}
\author{A. Varykhalov}\affiliation{BESSY GmbH, Albert-Einstein-Strasse 15, 12489 Berlin, Germany}
\author{L. Patthey}\affiliation{Paul Scherrer Institute, Swiss Light Source, CH-5232 Villigen, Switzerland}
\author{M. Shi}\affiliation{Paul Scherrer Institute, Swiss Light Source, CH-5232 Villigen, Switzerland}
\author{J.B. Goedkoop}\affiliation{Van der Waals-Zeeman Institute, University of Amsterdam, 1018XE Amsterdam, The Netherlands}
\author{M.S. Golden}\affiliation{Van der Waals-Zeeman Institute, University of Amsterdam, 1018XE Amsterdam, The Netherlands}\email{M.S.Golden@uva.nl}
\maketitle

\section{Dependence of the transport properties on the dimensionality}

\begin{figure*}[!b]
\includegraphics[width=1.0\columnwidth]{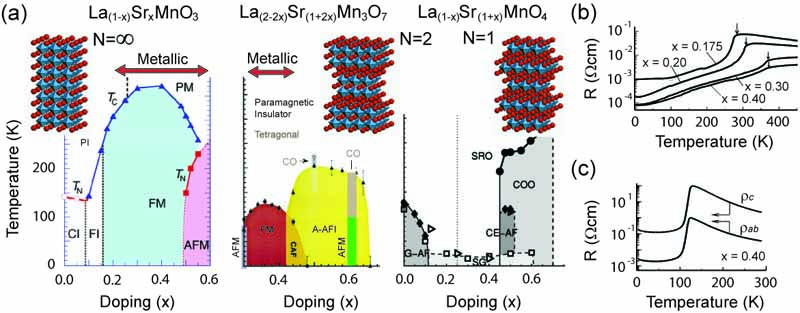}
\caption{\label{S2} Magnetic and transport properties of the LSMO Ruddlesden-Popper series. (a) Magnetic phase diagram of (La,Sr)$_{N+1}$Mn$_N$O$_{3N+1}$ as a function of Sr doping for the perovskite analogues N = 1 (left, after Ref.~\cite{tokura}), the bi-layered N = 2 (middle, after Ref.~\cite{ling, li}) and single layered N = 1 (right, after Ref.~\cite{Larochelle}). (b) Resistivity as function of temperature for various metallic compositions of perovskite LSMO (N = $\infty$, after Ref.~\cite{urushiba}). (c) Resistivity as function of temperature for bi-layered LSMO with x = 0.4 (from Ref.~\cite{kimura}).}
\end{figure*}

Figures~S\ref{S2}a-c show the phase diagrams taken from literature of (La,Sr)$_{N+1}$Mn$_{N}$O$_{3N+1}$, for N = 1 \cite{tokura}, 2 \cite{ling, li} and $\infty$ \cite{Larochelle}. Comparison of the diagrams clearly illustrates the strong dependence of the degree of apparent metallic character on the number of stacked MnO$_2$ planes, N. Whereas the cubic compound is truly metallic over a wide range of Sr doping content, the N = 1 composition is not metallic for any doping concentration. The N = 2 bilayer material is somewhere in between these two extremes, displaying metallic transport, but with an absolute value of resistivity in excess of the Mott maximum. To stress this latter point, Fig.~S\ref{S2}b and Fig.~S\ref{S2}c show the temperature dependence of resistivity for various doping levels of the cubic compound \cite{urushiba} and the x = 0.4 doping concentration of the bilayered LSMO material \cite{kimura}, respectively. As can be seen, the absolute value of resistivity of the cubic composition on the border of the insulating phase (x = 0.175) is still considerably less than the `metallic' x = 0.4 bilayered material. This is also apparent from optical conductivity measurements, where Drude weight is absent in the bilayered material \cite{optics_lsmo1, optics_lsmo2}, while it is clearly present in the cubic compound \cite{optics_cubic}.\newpage

\section{Sample and cleavage surface quality of the bilayer LSMO crystals}

\begin{figure*}[!b]
\includegraphics[width=1.0\columnwidth]{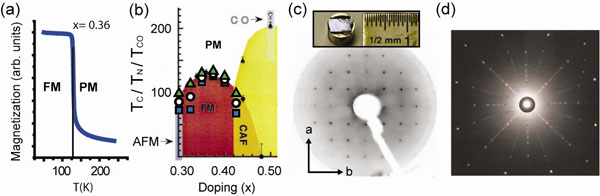}
\caption{\label{S1} Sample and cleavage surface quality of the bilayer LSMO crystals. (a) Magnetization versus temperature for an x = 0.36 sample measured using SQUID magnetometry. Data taken after zero field cooling, with an external field B = 100 G $||$ c. The sample shows a sharp transition from a paramagnetic (PM) to a ferromagnetic (FM) state at T$_{C}$ = 130 K with a total transition width of less than 5 K. (b) Magnetic transition temperatures as measured using SQUID magnetometry of LSMO with x = 0.30 (T$_{N}$) and 0.325 $\geq$ x $\geq$ 0.425 (T$_{C}$), plotted on top of the magnetic phase diagram, taken from Ref.~\cite{ling}. Depicted are the onset, midpoint and endpoint temperatures of the transition (triangles, circles and squares). (c) LEED image of a typical bilayer LSMO sample, E$_0$ = 400 eV, showing a very clear tetragonal pattern, without any signs of a 2D structural reconstruction. The inset shows a cleaved crystal on top of a cleavage post, with a smooth and mirror-like surface over millimeters. (d) Typical Laue image from our bilayer manganite crystals, indicative of the high quality and single crystallinity of the samples.}
\end{figure*}

The samples measured throughout this study were grown using the optical floating zone technique in both Oxford (x = 0.30, 0.325, 0.35, 0.375, 0.40, 0.425, 0.475) \cite{boothroyd1,boothroyd2} and Amsterdam (x = 0.30, 0.36, 0.40). Electron probe micro-analysis measurements on selected LSMO samples revealed that the composition of the single crystals was within a few percent of the nominal doping concentration and did not vary substantially across the crystal surface. All measured crystals were first characterised by magnetometry to ensure the sample has a sharp magnetic transition. Figure~S\ref{S1}a shows a typical magnetization versus temperature trace, showing a sharp transition into the low temperature ferromagnetic phase. Figure~S\ref{S1}b shows the phase diagram constructed from the transition temperatures of all doping concentrations used in this study, plotted on top of a phase diagram taken from literature \cite{ling}. The compositions used span the entire Sr doping regime where the colossal magnetoresistant effect occurs. Only crystals were selected which have a sharp magnetic transition, and are optically flat and shiny (see the inset of Fig.~S\ref{S1}c). Each cleavage surface measured with STM/S or ARPES has been investigated with low energy electron diffraction (LEED). Very sharp diffraction patterns were obtained of unreconstructed (1$\times$1) tetragonal symmetry. The single crystallinity of the samples was confirmed by Laue measurements, of which Fig.~S\ref{S1}d is a typical example. All of the above show the investigated single crystals and their cleavage surfaces to be of excellent quality \cite{PRBfig}.\newpage

\section{Origin of the QP-peaked areas}
\subsection{Fermi surface maps and T-dependent ARPES data}

\begin{figure*}[!b]
\includegraphics[width=1.0\columnwidth]{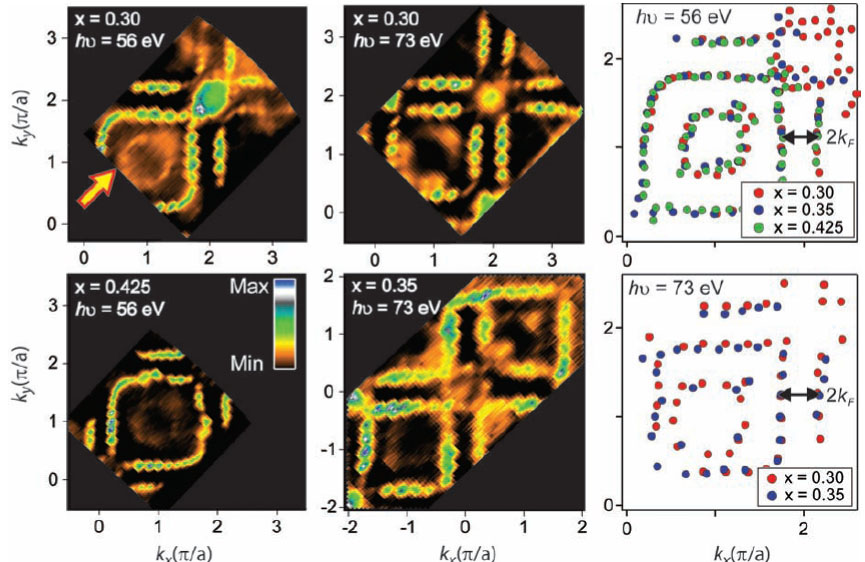}
\caption{\label{S7} Large k$_{||}$--area Fermi surface maps (signal integrated over E$_F$ $\pm$ 10 meV) for bilayer LSMO with x = 0.30, 0.35 and 0.425 taken at T = 20 K with h$\nu$ = 56 and 76 eV. The point density perpendicular to the analyzer slit is $\approx$ 0.08$\pi$/a (for h$\nu$ = 56). The large arrow in the top-left FS map indicates an unexpected electron-like Fermi surface pocket around the ($\pi$, $\pi$) point of k-space. The displayed data have been interpolated with two extra slices along the angular scanning direction (i.e. along the (0, 0) -- ($\pi$, $\pi$) direction). The two right-most panels show the fitted positions of the identifiable E$_{F}$ crossings in the raw FS maps for both photon energies. Results for the various doping level are overlain to allow for easy comparison. The error bars in the fitted positions are smaller than the symbol size.}
\end{figure*}

\begin{figure*}[!b]
\includegraphics[width=1.0\columnwidth]{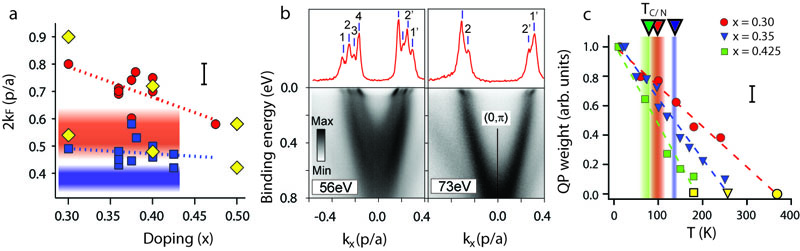}
\caption{\label{S3} Doping and temperature dependence of the electronic structure. (a) Doping dependence of the $2k_{F}$--values from $I(E,k)$--images recorded around the ($\pi$,0)--point of $k$--space (at $T=20$~K). Round and square symbols: $2k_{F}$--values for the inner and outer band of non--QP--peaked spectra. Shaded red and blue area: scatter of $2k_{F}$--values of the nominally bonding ($h\nu=73$~eV) and anti-bonding ($h\nu=56$~eV) bands from the QP--peaked regions of the cleaves, respectively. Yellow diamonds: $2k_{F}$--values predicted from density functional calculations \cite{LDA}. (b) $I(E,k)$--images at the ($\pi$,0)--point ($T=20$~K, $h\nu=56$ and 73~eV) showing as many as four different QP--peaked Fermi crossings, which are clearly identified in the $I(k)$--curves at $E_F$, shown as an inset to the panels. (c) Temperature dependence of the normalized QP--peaked spectral weight (integrated over $E_F\pm75$~meV) for three different doping levels. The respective bulk T$_{C}$'s of the samples are indicated.}
\end{figure*}

For a considerable proportion of all samples investigated, ARPES energy momentum maps spanning at least one Brillouin zone have been measured. Upon constructing a Fermi surface by fitting peaks in the momentum distribution curves at the Fermi energy, $k_{F}$-values can be extracted for each band crossing the Fermi level and compared to the values expected for the particular doping concentration. Representative Fermi surface maps for the QP-peaked region of cleaves of three different doping concentrations are shown in Fig.~S\ref{S7}. A similar analysis can be performed on the non-peaked regions of the sample, constructing a `ghost' Fermi surface by integrating the first 100 meV of the spectra. Figure~S\ref{S3}a shows the doping dependence of $k_{F}$-values of both peaked and non-peaked regions.

Immediately it becomes clear that the measured $k_{F}$-values for the peaked regions, aside from being - in some cases - too many in number, do not match with those expected from band theory \cite{LDA}. Moreover, there is no significant change in the various $k_{F}$-values as a function of doping. The highest and lowest doped compositions shown in Fig.~S\ref{S7} differ 12.5\% in hole doping, thus an equal change in Fermi surface area is expected between x = 0.30 and x = 0.35 and between x = 0.35 and x= 425. From LDA calculations \cite{LDA}, the 2k$_F$-values at ($\pi$, 0) for the anti-bonding and bonding band are expected to change between the two extreme doping levels considered from 0.54$\pi$/a and 0.90$\pi$/a to 0.48$\pi$/a and 0.65$\pi$/a, respectively. The right-most panels of Fig.~S\ref{S7} show the fitted k positions from the FS maps of the three doping levels overlain on top of each other. Noticeably, the data points for the three samples fall almost exactly on top of each other. This means that their Fermi surface areas are almost the same, despite the significant change in nominal hole concentration. The respective 2k$_{F}$ values at the ($\pi$, 0)-point are $\approx$ 0.46$\pi$/a and $\approx$ 0.36$\pi$/a for the h$\nu$ = 73 (nominally bonding band) and 56 eV (nominally anti-bonding band) data, respectively. Compared to the predictions from density functional theory, this would correspond to a doping level of x $\approx$ 0.50, thus a hole count some 10 to 20\% higher than nominal for the doping levels discussed here. Important to note is that a doping level of 50\% corresponds to an insulating, antiferromagnetically ordered low temperature phase of LSMO, a phase which is not expected to support spectral weight at the Fermi level. Over many measured cleavage surfaces of different doping levels, some scatter (about 10\%) has been observed in these respective 2k$_{F}$--values of the anti-bonding and bonding band QP peaks, but no systematic change in the area of the Fermi surface that scales with the hole doping is evident in our substantial collection of high resolution ARPES data. The $k$-values for the non-peaked regions on the other hand correspond nicely with calculations (within the uncertainty introduced by the very low spectral weight at E$_F$), and follow the expected trend as a function of doping concentration (see the blue and red symbols in Fig.~S4a).

Although often two peaks can be identified on the quasi--particle peaked areas, on multiple occasions not only single or double bands have been observed, but also triple and quadruple bands (see Fig.~S4b). The existence of two peaks can readily be explained by the presence of a bonding and anti-bonding band due to the double layer of manganese oxide octahedra. However, it requires a lot more effort to explain more than two bands for a perfectly bilayer crystal. The symmetric intensity distribution of the peaks at positive and negative k$_{x}$ and the switching of intensity between the spectra taken with different photon energy (top left panels of Fig.~S\ref{S7}), make it highly unlikely that the multitude of peaks is caused by simultaneous measurement of several crystallites with different doping level or surface angles with respect to each other. Without having a microscopic explanation for such multi-peaked data ready and waiting, here we simply show such data to make the reader aware of the intrinsic variability in the spectra from the QP-peaked regions of the sample surfaces. 

In addition to comparison with calculations, the temperature dependence of the quasi-particle peaks can also give clues as to the origin of the QP-peaks. If the peaked spectra indeed are a true signature of the bilayer manganite, the presence of the peaks should be highly correlated with the transition from the insulating to poor metallic state. For several doping concentrations the temperature dependence of the quasi-particle weight has been examined, the result of which is plotted in Fig.~S\ref{S3}c. Clearly, for none of the doping concentrations investigated do the peaks at the Fermi level vanish at T$_{C}$. Instead, quasi-particle spectral weight is in some cases even present to temperatures nearly as high as the metal to insulator transition of the cubic compound.

\begin{table*}[htb]
	\begin{center}
\begin{tabular}{ | c | c || c | c || c | c || c | c ||}
  	\hline
    doping (x) & QP abundance & doping (x) & QP abundance & doping (x) & QP abundance & doping (x) & QP abundance\\ \hline \hline
    0.30 & 3\% & 0.35 & 0\% & 0.36 & 1\% & 0.38 & 4\% \\ \hline
		0.30 & 5\% & 0.35 & 0\% & 0.36 & 0\% & 0.40 & 0\% \\ \hline
		0.30 & 5\% & 0.36 & 4\% & 0.36 & 0\% & 0.40 & 0\% \\ \hline
		0.30 & 4\% & 0.36 & 3\% & 0.36 & 0\% & 0.40 & 0\% \\ \hline
		0.30 & 0\% & 0.36 & 0\% & 0.36 & 0\% & 0.40 & 0\% \\ \hline
		0.30 & 4\% & 0.36 & 3\% & 0.36 & 0\% & 0.425 & 4\% \\ \hline
		0.30 & 5\% & 0.36 & 5\% & 0.36 & 0\% & 0.425 & 5\% \\ \hline
		0.30 & 0\% & 0.36 & 4\% & 0.36 & 0\% & 0.475 & 0\% \\ \hline
		0.30 & 0\% & 0.36 & 5\% & 0.36 & 0\% & 0.475 & 5\% \\ \hline
		0.325 & 0\% & 0.36 & 1\% & 0.375 & 10\% &  &  \\ \hline
		0.35 & 0\% & 0.36 & 3\% & 0.375 & 5\% &  &  \\ \hline
		\hline
\end{tabular}
\label{S4}\caption{Samples measured and the estimated area-percentage supporting quasi--particle peaks per sample, as determined by scanning the surface in ARPES using 100 micrometer size steps. A few percent at most of the total area of a sample supports sharply peaked features in k-space. Only samples are shown which have been spatially mapped in detail.}
\end{center}	
\end{table*}

Table~SI shows a list of samples measured in this study. Doping concentrations spanning the entire phase space in which the CMR effect occurs have been measured, with multiple samples for each doping concentration. Since not all samples measured in this study have been spatially mapped in detail, the list is not exhaustive. At most a few percent of the area of each sample supports quasi--particle peaked areas in ARPES, whereas the remaining sample surface shows a (pseudo-)gapped spectrum, the latter being readily unified with data from bulk (optics, transport) and surface sensitive spatially resolving techniques (STS) cited throughout the main text of this paper.

\subsection{$\sqrt{2}$x$\sqrt{2}$ in-plane symmetry for the QP-features in ARPES}

\begin{figure*}[!b]
\includegraphics[width=1.0\columnwidth]{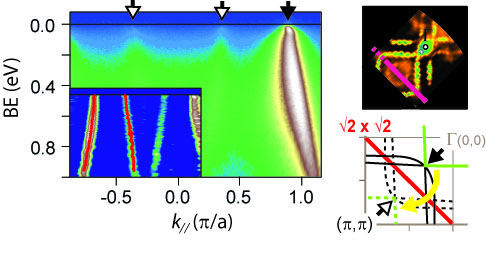}
\caption{\label{S5} The QP-peaked areas of a cleavage surface show a $\sqrt{2}$x$\sqrt{2}$ reconstruction. $I(E, k)$ image obtained from an x = 0.375 sample (h$\nu$ = 56 eV, T = 20 K) showing a copy (indicated with open triangles) of the main band (indicated with a closed triangle). The $I(E, k)$ image was recorded along the (0, 0)--($\pi$, $\pi$) direction in k-space, as indicated with a thick purple line in the small FS map (top right). The inset to the $I(E, k)$ image shows a second derivative of a similar spectrum for x=0.30, with the red lines showing the results of a fit to the peaks in the I(k) curves for the main band and one branch of the diffraction replica. The bottom right schematic FS illustrates how folding of the bands due to the presence of a $\sqrt{2}$x$\sqrt{2}$ reconstructed (smaller) Brillouin zone can lead to the observed band with its band bottom at ($\pi$, $\pi$).}
\end{figure*}

Another discrepancy between the measured Fermi surfaces from the QP-peaked regions and the predicted band structure for bilayer LSMO is a feature visible in all FS maps, for example that marked with a yellow arrow in Fig.~S\ref{S7}. A small circular pocket can be identified, which is electron-like around ($\pi$,$\pi$) (and is thus markedly different from the band structure features discussed throughout: these all being hole-like around the ($\pi$,$\pi$) point). The dispersion of the additional electron pocket can be seen in the E(k)-image cutting through k-space along the ($\pi$,$\pi$) - (-$\pi$,-$\pi$) direction depicted in Fig.~S\ref{S5}. The most intense feature in the spectrum is the main band, either the d$_{z^2-r^2}$ or the anti-bonding or bonding d$_{x^2-y^2}$ band that has its band bottom below BE = 1 eV at (0,0). Two other Fermi crossings, forming the extra electron pocket, can be seen in the image (indicated with open triangles).

Taking a second derivative of the spectrum (in k) enhances the contrast between the main band and these features. Such a second derivative spectrum is depicted in the inset to the large E(k)-image in Fig.~S\ref{S5}. The fitted band dispersions are plotted on top of the derivative spectrum and show that the extra features mirror the main band dispersion perfectly. Such a copy of the main band(s) can only occur if these are folded back around a smaller Brillouin zone boundary, resulting from a larger periodicity in real space than the tetragonal unit cell. From the observed dispersion of the diffraction
copies, we deduce that the reconstruction of the tetragonal unit cell is most likely of $\sqrt{2}\times\sqrt{2}$ origin and the diffraction copy is that of the d$_{z^2-r^2}$ hole pocket originating from around (0,0), as illustrated in the bottom right inset to Fig.~S\ref{S5}. 

\begin{figure}[!t]
\includegraphics[width=0.8\columnwidth]{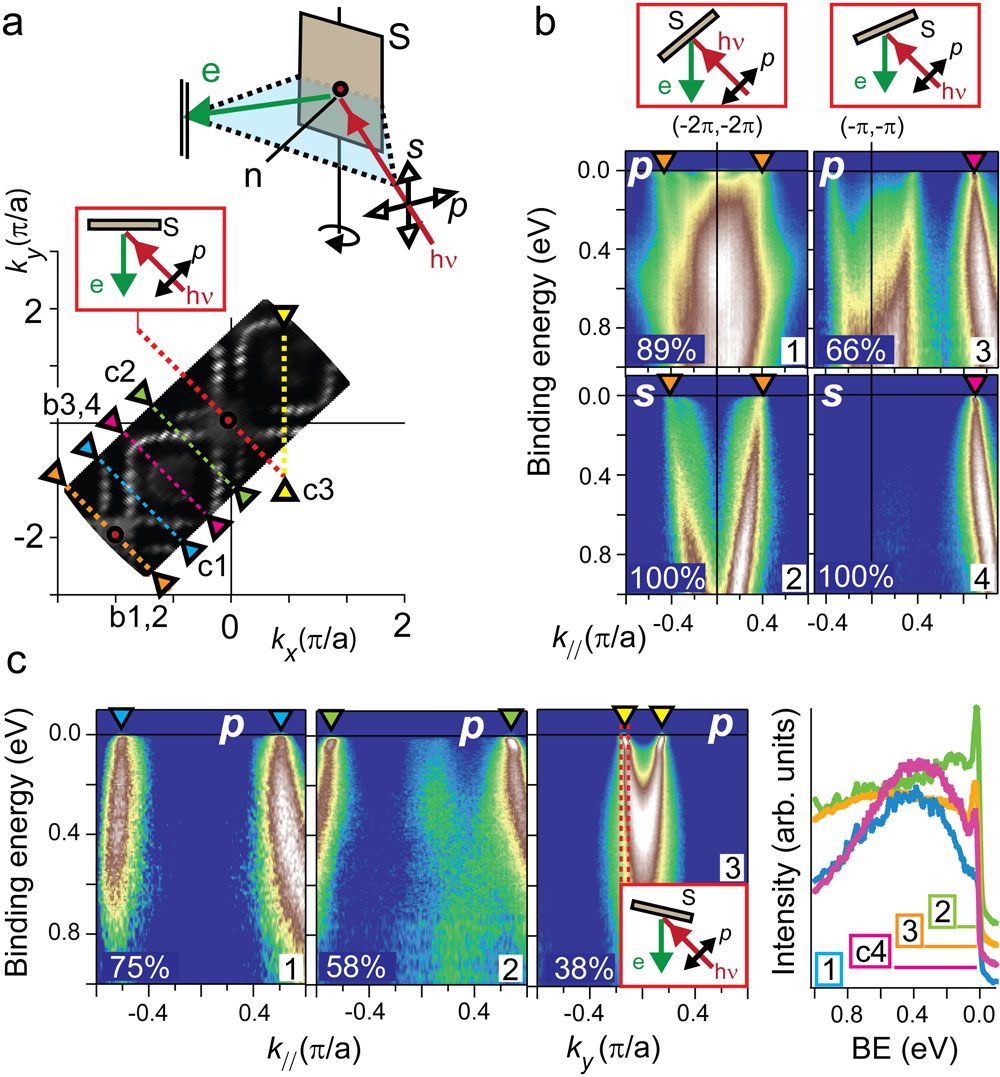}
\caption{\label{SInfo_geom} QP--peaked ARPES data from bilayer LSMO ($x=0.35$, $T=20$~K) recorded using different measurement geometries and photon polarizations. (a) Top: sketch of the measurement geometry, whereby `n' is the surface normal and `e' the outgoing photoelectron beam on its way to the analyzer slit. The black double-headed arrows indicate the polarization of the synchrotron radiation: either perpendicular (s) or parallel (p) to the plane spanned by the incoming light, the outgoing photoelectrons and the sample normal. Panel (a) bottom: large area Fermi surface map of bilayer LSMO (p--polarized light) recorded by rotation of the sample around the axis indicated in the sketch. A schematic (plan view) of the measurement geometry is shown for data taken through the $(0,0)$ $k$--space location. Note that the in--plane component of the light polarization increases upon rotation of the sample so as to reach the $(-2\pi,-2\pi)$ point. The cuts made through the dataset that are shown in panels (b) and (c) are indicated using colour-coded triangles. (b) I$(E,k)$--images from two different Brillouin zone diagonal cuts in $k$--space through the main hole-like bands, measured both with p-- and s--polarization, including sketches of the measurement geometry. The in--plane component of the polarization of the incoming photons is marked on each I$(E,k)$ image as a percentage. (c) Three I$(E,k)$--images from $k$-space cuts through the `zone face' region of the Fermi surface, recorded with p--polarization. Again, the in--plane polarization component of the photon polarization is given, as well as a sketch of the measurement geometry for sub-panel c3. The right-most sub-panel shows I$(E)$--curves for the three I$(E,k)$--images in panel c and the one in panel b4, obtained by integration of a small $k$-range around $k_F$ (as indicated with the red dotted box in sub-panel c3) and offset for clarity.}
\end{figure}

From the above observation, it follows that QP-peaked regions of the bilayer LSMO cleavage surface do not posses a 1$\times$1 tetragonal unit cell. This is in sharp contrast with the bulk crystal structure of bilayer LSMO in the ferromagnetic, metallic low temperature phase, which is strictly tetragonal. In addition, a surface sensitive diffraction technique such as LEED (that is however spatially averaging over $\sim$ 1mm$^{2}$) yields a strictly 1$\times$1 tetragonal diffraction pattern with a unit cell of 3.8 \AA. Interestingly, no diffraction copies were ever observed on the non-QP peaked areas of a cleavage surface. Thus, observations of the $\sqrt{2}\times\sqrt{2}$ reconstructed crystal structure are confined to the QP-peaked spots only. Small area electron diffraction experiments in transmission electron microscopy studies on bilayered LSMO with x = 0.40 have shown that intergrowth inclusions have a $\sqrt{2}\times\sqrt{2}$ reconstructed unit cell, with respect to the tetragonal structure of the bilayer matrix \cite{seshadri}. This would be exactly the type of reconstruction that could explain the observed diffraction copies seen in our high resolution ARPES data.

\subsection{Effect of matrix elements: photon polarization and measurement geometry in ARPES}

The ARPES and STM data presented here are argued to make a convincing case for the fact that the true spectral response of bilayer LSMO in the CMR region of the phase diagram is not that of a metal: pseudo-gapped, without QP's and thus without a Fermi surface. When arriving at such a conclusion from ARPES data it is vital to consider the effects of the matrix elements, in which the choice of photon energy, polarization and experimental geometry can have a profound impact on the experimentally measured intensities. Already from the fact that the QP--peaked spectra (here reported to result from a small surface area of a typical bilayer LSMO crystal that displays stacking faults) have been reported to be most easily resolved using either a photon energy of $h\nu=56$ or $73$~eV \cite{dessau, dessau2, dejong}, can one infer that matrix elements play a role for these features. This may also hold for the measurement geometry and polarization of the incoming photons. In fact, the ARPES studies on bilayer LSMO reported in Refs \cite{mannella} and \cite{mannella2} have been taken with a measurement geometry in which the polarization vector of the incoming synchrotron radiation essentially lies in the sample surface plane. These two studies have shown bilayer LSMO with a doping level of $x=0.40$ to be a `nodal metal'.

Therefore, we need to examine just how robust the observation (or not) of the QP--peaks is for different experimental conditions. Fig.~S\ref{SInfo_geom} shows ARPES data from a QP--peaked portion of a crystal cleave with $x=0.35$ (i.e. well inside the CMR portion of the phase diagram). In panel (a), apart from a sketch of the experimental geometry, we show a large area Fermi surface (FS) map, covering multiple Brillouin zones. Due to the measurement geometry adopted, the angle between the sample surface normal (denoted n) and the incoming light changes while recording different $k$-space regions of the FS map. When using p--polarized light, the photon polarization vector lies for $\approx89$\% in the sample surface plane for data recorded around $(k_x,k_y)=(-2\pi,-2\pi)$. Moving toward a normal incidence geometry - i.e. for data recorded around $(0,0)$ - the polarization vector has equally large in- and out-of-plane contributions . For data taken at the $(\pi,\pi)$ point of $k$-space the vector has an out-of-plane contribution of 66\%. On the other hand, for the same geometry, the polarization vector is 100\% in--plane, irrespective of the probed area of $k$--space, for data recorded with s--polarized light. Hence, the data contained in the FS map shown in Fig.~S\ref{SInfo_geom}a, taken together with s-polarization, offer an ideal test of the robustness of the presence of QP--peaked against polarization and geometry induced matrix element effects. 

Panel b of Fig.~S\ref{SInfo_geom} shows I$(E,k)$ images that cut $k$--space around $(-\pi,-\pi)$ and $(-2\pi,-2\pi)$, taken with both s-- and p--polarized light. The amount of in--plane polarization varies between 100 and 66\%, as marked in the figure. Despite this, in all four cases the bands comprising the $X$--centered hole-like FS, here cut along the zone diagonal, clearly show QP--peaks at $E_F$ (indicated by colored triangles whose colour matches the symbols marking the k-space cuts in the Fermi surface map). 

Additionally, the cut shown in panel b1 shows two electron--like bands centered around $(-2\pi,-2\pi)$, one of which is very shallow and has its band bottom below a binding energy of $BE=50$~meV. Panel b3 shows a copy of these features centered around $(-\pi,-\pi)$, due to the decreased in-plane crystal symmetry of the stacking fault intergrowths, as has been discussed in connection with Fig.~S5. 

The same cut in $k$--space taken with s--polarized light (panel b2) shows a strong suppression of the two electron-like features due to the selection rules. This also true for the $(-\pi,-\pi)$-located diffraction copy of these features in panel b4. 

The `extinction' of the two $\Gamma$--centered electron pockets (and their diffraction copies) upon use of s--polarized light indicates that they have mainly $d_{z^2-r^2}$ character, in keeping with predictions from LDA calculations \cite{LDA,SanizLDA} . Due to the measurement geometry, the selection rules only apply strictly at center of the analyzer slit (i.e. the centre of the k-space axis of each I$(E,k)$ image), as only this point lies in the mirror plane defined by the incoming photons, the surface normal and the outgoing photoelectrons. This hinders an unambiguous determination of the orbital character of the bands comprising the $X$--centered hole pocket. 

Irrespective of the selection rules, the QP--peak spectral weight of the hole-like features is almost constant compared to the incoherent spectral weight at higher binding energies across all four panels of Fig.~S6b. In particular, the similarity between panels b3 and b4 shows that the zone--diagonal QP--peak is still robustly observed, despite a significantly difference in in--plane polarization for the two `nodal' cuts.

This observation also holds for the QP--peaks observed around the zone face regions of $k$--space, see Fig.~S\ref{SInfo_geom}c. In this panel, three I$(E,k)$--images are presented that have a decreasing in--plane contribution of the photon polarization. Cut c1, with the highest in--plane contribution, does show reduced QP spectral weight compared to the other two cuts. However, it still has significant spectral weight at $E_F$ and is thus far from being `pseudo-gapped'. This can also be seen in the I$(E)$--curves for $k\sim~k_F$ depicted in the rightmost panel of Fig.~S\ref{SInfo_geom}c, which also includes a zone-diagonal $k_F$ I$(E)$--trace for comparison.

This section shows that although the measurement geometry does play a role in the intensity of the QP--peaks at $E_F$, the observation of QP--peaks at the various $k$--locations on the Fermi surface is not restricted to a particular set of experimental conditions, but is restricted to particular - minority - locations on the sample surface from which signals from intergrowth patches are picked up in the experiment.
Also important is the fact that the lack of QP--peaks from the majority of the cleavage surface is not a particular result, but generally applicable, also for experimental geometries with a large in-plane component of the photon polarization, and for a wide range of photon energies.
The modest sensitivity of the data from both the non-QP--peaked and QP--peaked regions to experimental conditions only serves to strengthen our conclusion that the general spectral signature of bilayer LSMO in the ferromagnetic region of the phase diagram is that of a Fermi-surface-less non-metal, without coherent spectral weight at $E_F$, and that the QP-features and Fermi surfaces we observe originate from N$\neq$2 intergrowths.\newpage

\section{Pseudo--periodic textures seen with STM}

\begin{figure*}[!b]
\includegraphics[width=1.0\columnwidth]{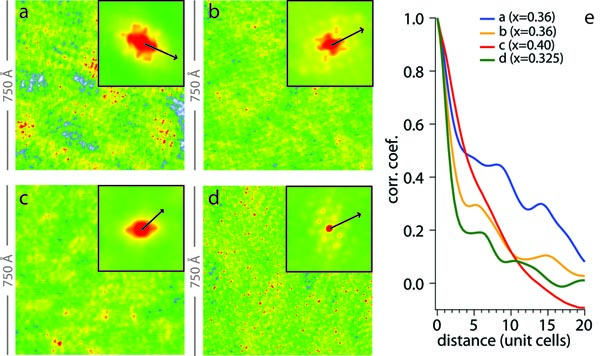}
\caption{\label{S6} (a)-(d) Pseudo--periodic texture observed in the topographic STM images from different Sr doping concentrations of bilayer LSMO. The setup conditions are similar to those in Fig.~1 of the main paper. The vertical corrugation of these semi-regular, long wavelength features is, at less than 1~\AA\, too small to be due to differing atomic layers of the crystal. Doping levels are given in the legend to panel (e), which shows line traces through the autocorrelation plots derived from the topographs. These traces show that the data support differing correlations of range between 5-10 unit cells. Each autocorrelation figure is shown as an inset to the corresponding STM topograph.
 }
\end{figure*}

The pseudo--periodic texture as seen with STM on the bilayer manganites has been observed on samples with various doping concentrations. Fig.~S\ref{S6} shows topographies for four different samples supporting such textures. Auto-correlation traces of these four topographs are shown in Fig.~S\ref{S6}e, in all cases having correlation peaks at distances ranging from $\sim$ 5-10 unit cells. There is no simple trend in these characteristic distances as a function of doping, and nor are the periodic structures pinned to the real-space lattice vectors.